# Improving 2D-ness to enhance thermopower in oxide superlattices

**Dongwon Shin[1,#], Inseo Kim[2,#], Min-Su Kim[3], Yu-Qiao Zhang[4], Woo Tack Lim[1], Si-Young Choi[3,5,6], Minseok Choi[2,*], Hiromichi Ohta[7,*], and Woo Seok Choi[1,*]**

[1] Department of Physics, Sungkyunkwan University, Suwon 16419, Republic of Korea
[2] Department of Physics, Inha University, Incheon 22212, Republic of Korea
[3] Department of Materials Science and Engineering, Pohang University of Science and Technology, Pohang 37673, Republic of Korea
[4] School of Chemistry and Chemical Engineering, Jiangsu University, Zhenjiang 212013, China
[5] Department of Semiconductor Engineering, Pohang University of Science and Technology, Pohang 37673, Republic of Korea
[6] Center of Van der Waals Quantum Solids, Institute for Basic Science (IBS), Pohang 37673, Republic of Korea
[7] Research Institute for Electronic Science, Hokkaido University, Sapporo 001-0020, Japan

*Author to whom any correspondence should be addressed.

E-mail: hiromichi.ohta@es.hokudai.ac.jp, minseok.choi@inha.ac.kr, and choiws@skku.edu

## Abstract

The thermoelectric performance of a material is determined by the fundamental transport dynamics of itinerant charge carriers and their interactions with the environment. For two-dimensional oxide thermoelectrics, predominantly represented by doped $SrTiO_3$-based superlattices, reduced spatial dimensions and increased effective mass are known to enhance thermopower ($S$). However, because of their large effective Bohr radius resulting from their high dielectric constant, $SrTiO_3$-based systems have limitations in exhibiting the 2D characteristic. Here, we focus on $EuTiO_3$ as an alternative perovskite platform in which fractional $La_xEu_{1-x}TiO_3/EuTiO_3$ artificial superlattices demonstrate the improvement in 2D nature for the dimensionality-induced improvement of $S$. We observed a quasi-2D thermopower $S_{2D}$ of $-950$ μV K$^{-1}$ and $S_{2D}/S_{3D}$ of ~20 resulting from the improved 2D confinement. Thermopower measurements, combined with hybrid density functional theory calculations, show the enhanced $S$ originates from the confinement of Ti $3d_{xy}$-states within the $La_xEu_{1-x}TiO_3$ layers and the associated increase in the 2D density of states. In detail, a smaller effective Bohr radius and modified electronic band structures, in conjunction with the presence of the Eu $4f$-states in $EuTiO_3$ which modified the local electronic potential and strengthened the spatial confinement of Ti $3d$-states. This approach to improving the dimensional confinement establishes a small effective Bohr radius and Eu $4f$-state assisted 2D confinement provides valuable insights into the design of high-performance applications in artificial oxide superlattices.

## 1. Introduction

Thermoelectric properties of a material are determined by the transport characteristics of itinerant charge carriers, which are influenced by their interactions with the surroundings. Thermopower generation, or the Seebeck effect, converts waste heat into useful electricity and is one of the most important practical technologies for sustainable energy generation [1-7]. The thermoelectric efficiency is quantified by a dimensionless figure of merit, $ZT = S^2\sigma T\kappa^{-1}$, where $S$, $\sigma$, $\kappa$, and $T$ are the thermopower or Seebeck coefficient, electrical conductivity, thermal conductivity, and absolute temperature, respectively. Unfortunately, conventional trade-off relations, such as the negative correlation between $S$ and $\sigma$ with respect to the carrier density ($n$), set up a conventional limitation in the improvement of $ZT$ values [3, 8]. Recent studies on thermoelectrics have focused on overcoming such limitations by manipulating the characteristics of the surrounding environment of the charge carriers with different strategies [9-15].

Perovskite transition metal oxides (TMOs) with strong electron correlation are promising model systems to attest the ramifications of the designed modulated surroundings for enhanced thermoelectric properties [16-21]. TMOs have advantages owing to their thermal and chemical stability, non-toxicity, and compatibility with high-temperature and application environments. In addition, TMOs exhibit similar energy scales of charge, lattice, spin, and orbital degrees of freedom [22], leading to a facile controllability based on epitaxial strain modulation, carrier concentration tuning, band engineering, and nanostructuring. This also facilitates the interaction between the itinerant charge carriers and their surroundings, which often forms dressed quasiparticles that do not behave according to the elementary transport theories of electrons.

Among the TMOs, electron-doped SrTiO₃ (STO) has been extensively studied as an *n*-type thermoelectric material (Table S1, Supplementary data). Compared to other *n*-type oxides, such as Al-doped ZnO [23], CaMnO₃ [24], In₂O₃ [25], electron-doped TiO₂ [26], and electron-doped STO [18, 27-30], the carrier concentration in electron-doped STO can be most easily controlled through La or Nb substitutional doping, enabling power factor ($PF = S^2\sigma$) optimization. La or Nb doping not only tunes the carrier concentration but also modifies the electronic band structure of STO, leading to increase in the effective mass of the electrons [28, 29, 31]. For example, thin films of La-doped STO (LSTO) with controlled elemental vacancies exhibit a correlation between the polaronic mass enhancement and $S$ [32].

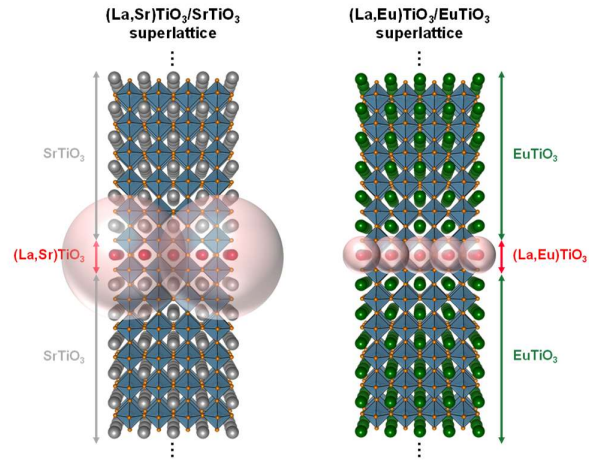

**Figure 1.** Improved 2D-ness in fractional LETO/ETO SLs via reduced $a_0$. The schematics illustrate LSTO/STO (left panel) and LETO/ETO SLs (right panel), where Sr, Eu, La, and O atoms are denoted as gray, green, red, and orange spheres, respectively, with blue TiO₆ octahedra. $a_0$ is shown as pink translucent spheres. Fractional LETO/ETO SLs host itinerant charge carriers with smaller $a_0$, compared to LSTO/STO SLs.

STO has also been primarily used as a model system to confine the itinerant charge carriers to a lower spatial dimension. This is a promising strategy to control the environment, as the modified electronic density of states (DOS) near the Fermi energy ($E_F$) might enhance both $S$ and $\sigma$ simultaneously [33, 34]. Particularly, STO-based superlattices (SLs) with La or Nb doping have been proposed to exhibit 2D transport characteristics with substantial enhancement of $S$ [21, 35-38]. However, the so-called 2D-ness of STO-based SLs have not been tested yet in comparison to other systems, which raises questions on the pure effect of the reduced dimensionality. Most importantly, the high dielectric constant ($\varepsilon$) of STO, which is often beneficial for electron mobility by screening the impurity potentials, has an adverse effect in confining the carriers to a lower dimension owing to the increased polarizability. In particular, the high $\varepsilon$ causes a large effective Bohr radius $a_0 = a_B\varepsilon/m^*$, where $a_B = 0.53$ Å is the hydrogenic Bohr radius and $m^*$ is the effective mass [39, 40], ill-defining the 2D characteristics, as schematically shown in the left panel of figure 1. To clarify the intrinsic distinctions between the conventional STO-based and our EuTiO$_3$ (ETO)-based SL systems, we provide a direct comparison of parameters in table 1 [21, 39-44]. This point-by-point summary includes key quantities such as the $\varepsilon$, $a_0$, and experimentally reported $S$ values. The comparison highlights that the ETO-based SLs offer a more favorable environment for 2D confinement due to their inherently smaller dielectric screening and shorter spatial extent of electronic wavefunctions. Hence, using TMOs with lower $\varepsilon$ might facilitate the 2D confinement, thereby promoting our understanding and the development of highly efficient low dimensional thermoelectric materials.

**Table 1.** Comparing intrinsic factors for 2D-ness of fractional LSTO/STO and LETO/ETO SLs.

| At 300 K | (La,Sr)TiO$_3$/SrTiO$_3$ superlattice | (La,Eu)TiO$_3$/EuTiO$_3$ superlattice (This study) |
|---|---|---|
| Lattice parameter of SrTiO$_3$ (EuTiO$_3$) | 3.905 Å | 3.905 Å |
| Dielectric constant $\varepsilon$ | 300 | 70 |
| Effective Bohr radius $a_0$ | 150 Å | 30 Å |
| $|S_{3D}|$ | 28 – 63 $\mu$V K$^{-1}$ | 18 – 443 $\mu$V K$^{-1}$ |
| $|S_{2D}|$ | 160 – 407 $\mu$V K$^{-1}$ | 229 – 907 $\mu$V K$^{-1}$ |
| $S_{2D}/S_{3D}$ | < 10 | < 20 |
| Orbital state near $E_F$ | – | Eu 4$f$-state |

In this study, we explore the various parameters of dimensionality and electronic correlation, offered by synthetic crystals comprising perovskite TMOs with low $\varepsilon$. We employed artificial oxide SLs consisting of ETO and LaTiO$_3$ (LTO) perovskite building blocks. Unlike prior approaches to rely on STO as the host material, our work represents the first direct comparative study between STO- and ETO-based systems with respect to their intrinsic ability to support 2D confinement. By introducing ETO, which possesses a fundamentally distinct electronic structure and short $a_0$, we go beyond material substitution and demonstrate how its physical parameters enable much stronger confinement, ultimately leading to greater $|S|$ enhancement. ETO exhibits a smaller $\varepsilon_{ETO}$ of ~70 at room temperature, compared to that of the STO ($\varepsilon_{STO} = $~300) in bulk phase [45, 46]. $a_0$ of STO and ETO are approximately 150 Å (with $\varepsilon = 300$ and $m^* = 1.0 \sim 1.1$) [29] and 30 Å (with $\varepsilon = 70$ and $m^* = 1.2$) [40], respectively, showing a five-fold difference between them, as schematically presented in figure 1. The reduced electronic wavefunction in the ETO-based SLs results in an enhanced 2D-ness (right panel) compared to the STO-based SLs (left panel). In addition to a similar Ti 3$d$-orbital character, ETO has Eu 4$f$-orbitals, which is located near $E_F$, in comparison to the STO electronic structure. Because of the distinct Eu 4$f$-orbital states [47, 48], ETO exhibited a large $|S|$ of 1,000 $\mu$V K$^{-1}$ in polycrystalline form (figure S1(a), Supplementary data) [41-43]. On the other hand, perovskite LTO was selected to mimic the LTO/STO SLs, in which the LTO layer is expected to provide and control the charge carriers through interfacial electronic reconstruction (figure S1(b), Supplementary data) [49-52]. By inserting one-unit-cell (u.c.)-thick La-doped ($x$) ETO (La$_x$Eu$_{1-x}$TiO$_3$, LETO) layers ($\delta$-doped layer) within the ETO matrix, i.e., comprising the fractional $\delta$-doped LETO/ETO SLs, we modulate the correlation effect of the 2D confined $d$-electrons. An approximate 20-fold enhancement of $|S|$ in the fractional LETO/ETO SLs was observed compared with the single films at $x = 0.6$. This maximum $S_{2D}/S_{3D}$ value exceeds that of the previously reported LSTO/STO SLs (~9) [21], which highlights the 2D confinement effect in TMOs with low $\varepsilon$.

## 2. Results and Discussion

### 2.1 Structural properties of LETO single films and fractional δ-doped LETO/ETO SLs

High-quality epitaxial LETO single films and fractional δ-doped LETO/ETO SLs with systematically modulated *x* were fabricated by pulsed laser epitaxy. We first show the lattice structures of LETO single films, which are crucial for comprehensively explaining the La-doping effect in the 3D case before approaching to the 2D case (figure 2(a)). Figure 2(b) and 2(c) show the X-ray diffraction (XRD) $\theta$–$2\theta$ scans and X-ray reflectivity (XRR) of the LETO single films grown on $(LaAlO_3)_{0.3}(Sr_2TaAlO_6)_{0.7}$ (LSAT) substrates, respectively, which demonstrate the high quality of the single-crystalline epitaxial thin films. Expanded XRD $\theta$–$2\theta$ scans and XRD reciprocal space mappings (RSM) further confirm the single phase and strain state of the epitaxial LETO single films, respectively. (figure S2, Supplementary data) Considering the distinctive strain state of $x = 1.0$ and the highly insulating nature of the $x = 0$ single films [43], we focus on the transport properties of epitaxial LETO single films with $x = 0.05, 0.1, 0.2, 0.4, 0.6$ and $0.8$. On the other hand, we also fabricated high-quality fractional LETO/ETO SLs with similar systematic modulation of *x* at the LETO layer (figure 2(d)), as in the LETO single films. XRD $\theta$–$2\theta$ scans and XRR results elucidate the atomically well-defined periodicities of the fractional LETO/ETO SLs, as shown in figure 2(e), 2(f) and figure S3(a) (Supplementary data). The fractional LETO/ETO SLs were fully strained to the LSAT substrates, as evidenced by the XRD RSMs near the (103) plane of the LSAT substrate (figure S3(b), Supplementary data). While the LETO single films and fractional LETO/ETO SLs are fully strained to the LSAT substrates, it is worth noting that strain may still play a role in modulating the electronic structure and transport properties, particularly if different substrates with varying lattice mismatches are employed. Our study did not explore substrate-dependent strain effects, therefore, the strain state is uniform and can be excluded as a variable affecting the thermoelectric behavior in this study.

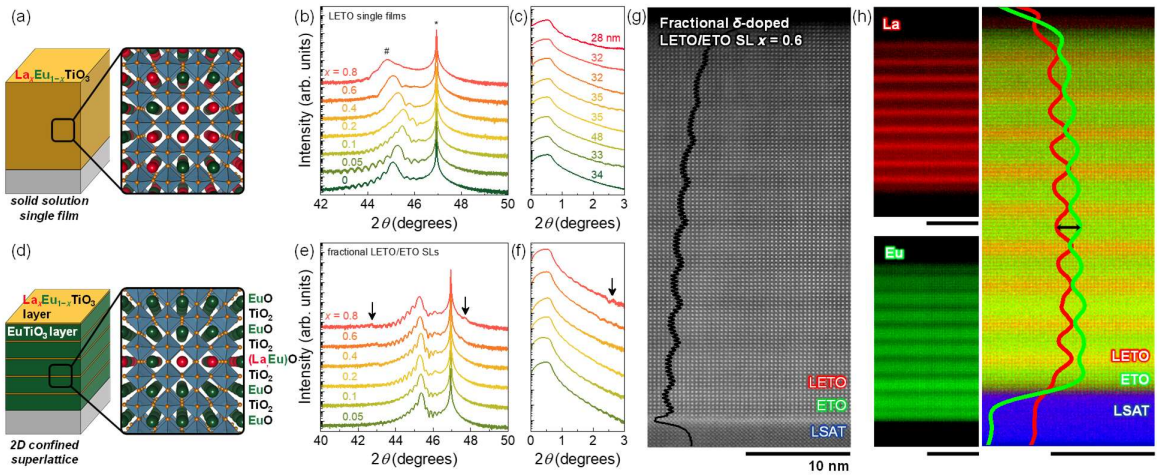

**Figure 2.** Atomic precision growth of epitaxial LETO single films and fractional LETO/ETO SLs. (a) Schematical representation of LETO single film, showing an expanded view of the atomic arrangements. La, Eu, and O are shown as red, green, and orange spheres, respectively, with blue $TiO_6$ octahedra. (b) XRD $\theta$-$2\theta$ scans near (002) Bragg planes of the epitaxial LETO single films (#) on (001)-oriented LSAT substrates planes (*), respectively. (c) XRR of the epitaxial LETO single films with various *x*. (d) Schematic representation of fractional LETO/ETO SL, showing the 2D confinement of the LETO layer within a single u.c with an expanded view of the atomic arrangements. (e) XRD $\theta$-$2\theta$ scans of the fractional LETO/ETO SLs confirming the SL Bragg peaks (labeled black arrows), representing the atomically controlled periodicity of the SLs around the (002) peak. (f) XRR of the fractional LETO/ETO SLs showing SLs periodic oscillations. (g) Cross-sectional *Z*-contrast HAADF-STEM images for fractional LETO/ETO SL $x = 0.6$. The scale (black bars) shown in all images of the SL is 10 nm. (h) The EDS identifies all the layers in the SL on the atomic scale with slight cation intermixing at the interface. The chemical formulas of the component materials in the SL are color-coded to correspond to the atomic-resolution EDS image on the left (La, red; Eu, green). The intensity difference between the Eu and La atoms additionally confirmed the artificial stacking of SL.

## 2.2 Microstructural analyses in fractional δ-doped LETO/ETO SLs

The 2D nature of the fractional δ-doped LETO/ETO SLs is further clarified from the atomic structures. We used high-angle annular dark-field scanning transmission electron microscopy (HAADF-STEM) and energy-dispersive X-ray spectroscopy (EDS) as shown in figure 2(g) and 2(h), respectively, for fractional LETO/ETO SL at $x = 0.6$. The intensity oscillations of each atomic layer in figure 2(g) indicate that the fractional SL comprises a total of 11-u.c.-thick supercell, as designed and confirmed using XRD. In addition, EDS mapping was conducted for the elemental mapping of the La (red), Eu (green) and Ta (blue) atoms to identify the periodic structure of the fractional LETO/ETO SL (figure 2(h)). We observed that the EDS mapping presents a thicker LETO layer (~3 u.c.) and a thinner ETO layer (~8 u.c.) than the expected structure (1 u.c. of the LETO layer and 10 u.c. of the ETO layers). Even considering the potential step and terrace structure of the substrate, this diffuse interface might indicate the potential intermixing of Eu and La ions, at least compared to the cases of Sr and La ions in the LSTO/STO SLs [56, 57]. The intermixing may be attributed to the high diffusivity of Eu ions, which is smaller than that of Sr ions. Similar phenomena were observed at the interfaces of multilayer oxide thin films containing Eu ions, for example, LaAlO$_3$/ETO on STO substrates and LaAlO$_3$/ETO on Sr$_{0.99}$Ca$_{0.01}$TiO$_3$ substrates, from similar EDS maps during STEM [58, 59]. Similar structural coherency with a designed supercell thickness of 11 u.c. is shown in $x = 0.1$ and 1 SLs, although the contrast is now much weaker at $x = 0.1$ SL, as anticipated (figure S4 and S5, Supplementary data). Although defects and interfacial effects can affect transport in oxide heterostructures, it is important to note that both the LETO single films and the fractional LETO/ETO SLs were fabricated under the same growth conditions, including substrate type, temperature, pressure, and deposition rate. Therefore, the type and concentration of unintended defects (such as oxygen vacancies or cation mixing) are comparable across the different sample geometries. Consequently, any transport contributions arising from such defects are likely to cancel out in a relative comparison. Moreover, because all transport measurements, including thermopower, were performed along the in-plane direction, the influence of interface scattering (typically more prominent in out-of-plane transport) can be minimal in our study.

## 2.3 Thermoelectric performance with the 2D confinement effect in low-dimensional oxide

Both $|S|$ of LETO single films and fractional LETO/ETO SLs are extracted from linear fitting of the $\Delta V$-$\Delta T$ curves (figure S6, Supplementary data). To compare the results on an equal footing, $|S|$ values at 300 K are reported. We demonstrate the overall thermoelectric properties of epitaxial LETO single films (3D case) and STO-based samples from previous studies by showing the universal $|S_{3D}|-\log n$ relation. $n$ was extracted from transport measurements. (figure S7, Supplementary data). The universality confirms that the 3D charge transport characteristics are mostly determined by the Ti $3d$-orbitals of both LETO and STO-based samples, including the LSTO thin films. $S$ of the semiconductors can be explained by the relaxation time approximation of the Boltzmann's equation (see Section S5 in the Supplementary data) [29], which is represented by the blue dotted line in figure 3(a). The equation can be simplified as $|S_{3D}| = -k_B/e \times \ln 10 \times A \times (\log n + B)$, where $k_B$ is the Boltzmann constant, $e$ is the electron charge, $A$ and $B$ are parameters dependent on the types of materials and the electronic band structures, respectively [60]. All the $|S_{3D}|$ values of the LETO single films (colored open circles) and previously reported electron-doped STO samples (black open squares) [21, 28, 30, 35] universally align with the theoretical expectations of $|S_{3D}| = (-198 \text{ μV K}^{-1})$ $\times (\log n + B)$. Now, one can expect a deviation from this universal curve, as the DOS near $E_F$ changed with the 2D confinement [35]. In the fractional LETO/ETO SLs, direct electrical transport measurements except for $\sigma$ were not feasible in our current experimental setup (figure S8, Supplementary data). The transport signals were below the reliable detection limit, particularly at low doping levels where the system becomes less metallic. To analyze $|S|$ in fractional LETO/ETO SLs with the absence of direct transport data, we adopted an alternative approach based on nominal $n$, estimated from the La doping ratio $x$ [61]. Under the assumption that each La$^{3+}$ substituting Eu$^{2+}$ introduces one conduction electron, the $n$ is expected to scale linearly with $x$. Using these nominal $n$ values, we constructed $|S|$ - log $n$ of fractional LETO/ETO SLs, as shown in figure 3(a). As noted above, such approach might suffer from cation intermixing at the interfaces, as evidenced by STEM-EDS, resulting in potential change in the Eu valence state. Oxygen vacancies and other defects may further affect $n$. Additionally, strong electronic correlations or localization effects could trap carriers, making the nominal value rather distinct from actual $n$. Nevertheless, growth conditions were identical across the SLs, so extrinsic contributions such as oxygen vacancies, Eu valence variations, and cation intermixing, if any, are expected to be comparable between the samples. Hence, while the nominal $n$ may deviate from the actual $n$, this uniformity minimizes its impact on relative trends, and the observed doping-dependent evolution of $|S|$ across the fractional SLs provides compelling evidence with dimensional confinement. Using nominal $n$ values, $|S_{2D}|$ exhibits consistency with general trends of Boltzmann transport theory observed in semiconductors but aligns with the theoretical expectations of $|S_{2D}| = (-706 \text{ μV K}^{-1}) \times (\log n + B)$. In the absence of direct transport measurements, this approach offers empirical presentation of the enhancement of $|S|$ in fractional SLs due to the 2D confinement.

A strong 2D enhancement of |S| was observed in fractional LETO/ETO SLs, owing to the confinement of *d*-electron within the LETO layer. In figure 3(b), we show the systematic modulation of |$S_{3D}$| for epitaxial LETO single films (open circles) The |$S_{3D}$| value for *x* = 0.05 LETO single film (dark green open circle) was 450 μV K⁻¹, which gradually decreased with increasing *x*, to 18 μV K⁻¹ for *x* = 0.6. Interestingly, |$S_{3D}$| slightly increased for *x* = 0.8 LETO single film because of the $m^*$ enhancement (figure S7 and table S2, Supplementary data), as the system approached the Mott boundary located near *x* = 0.8 (see detailed explanation in section S5, Supplementary data). In the 2D case, the |$S_{2D}$| value for *x* = 0.1 of fractional LETO/ETO SL was 950 μV K⁻¹, which slowly decreased to 229 μV K⁻¹, for the *x* = 0.8 SL. The overall qualitative behavior of |$S_{2D}$| as a function of *x* was similar to that of the LSTO/STO SLs (figure S9, Supplementary data) [21, 44]. A comprehensive assessment of thermoelectric performance would ideally include the substrate contribution. However, because of the substrate, which is ~10⁴ times thicker than the nanometer-scale film, reliable extraction of in-plane *κ* of the film is not possible. We therefore focus on demonstrating the 2D confinement driven enhancement of |S|.

As expected, the 2D enhancement effect of |S| in the LETO system was substantial, more than twice as large as in the LSTO system. Figure 3(c) shows the values of $S_{2D}/S_{3D}$ obtained by comparing *S* of the SLs (2D) with that of a single film (3D). $S_{2D}/S_{3D}$ increased as *x* increased to 0.6, reaching almost 20. Along with the strong correlations responsible for the significant |S| of the single films, the reduced dimensionality induces further improved |S|. In particular, 2D confinement caused anisotropy in the band structure, which further strengthened with *x*, as the chemical contrast of the *δ*-doped layer with respect to the ETO layer increased. The anisotropic band structure resulted in a 2D enhancement of |S|. When *x* = 0.8, $S_{2D}/S_{3D}$ decreased owing to the modified electronic structure near the Mott boundary, as |S| of a single LETO film increased when *x* = 0.8, which deviates from the trend for *x* < 0.8. Similar behavior has been reported in La-doped STO thin films [44]. Interestingly, a similar qualitative behavior of $S_{2D}/S_{3D}$ is demonstrated for the LSTO system, however, the fractional LETO/ETO SLs exhibited twice the 2D enhancement compared to that of the LSTO/STO SLs [21, 44]. This observation is even more surprising, considering the apparent chemical diffusion (~3 u.c.) recorded by electron microscopy at the LETO/ETO interfaces. Despite ~3 u.c. intermixing at the LETO/ETO interfaces, the difference in electronic contribution, set by the $a_0$ (ETO ~ 30 Å and STO ~ 150 Å) in table 1, far exceeds the difference in chemical intermixing (ETO ~ 12 Å and STO ~ 4 Å). Thus, |$S_{2D}$| is dominated by electronic wavefunction confinement rather than chemical diffusion.

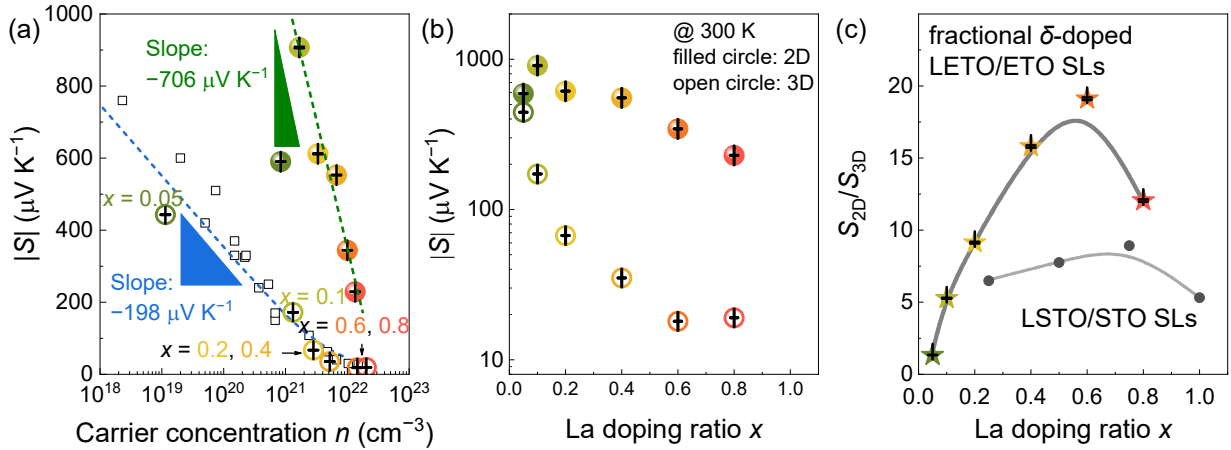

**Figure 3.** |S| enhancement in the epitaxial LETO single films and the fractional LETO/ETO SLs. (a) |S|–log *n* plots at 300 K for the electron-doped STO samples (previous reports; black open squares) and the experimentally obtained epitaxial LETO single films (this work; colored open circles). The slope of |S| is −198 μV K⁻¹, which is universally well-fitted in the case of the electron-doped STO samples and epitaxial LETO single films owing to similar conduction band of Ti 3*d*-orbital for both STO- and ETO-based samples. In fractional SLs, the slope of |S| is −706 μV K⁻¹ with respect to nominal *n* value shows 3.5 times larger due to the 2D confinement. (b) Experimentally measured |S| as a function of *x* for the epitaxial LETO single films (3D; opened circles) and fractional LETO/ETO SLs (2D; filled circles). The fractional SLs show highly enhanced |S|, mainly because of the 2D confinement of the *d*-electrons. (c) *x*-dependence of $S_{2D}/S_{3D}$ in thermoelectric oxide heterostructures. $S_{2D}/S_{3D}$ in the fractional LETO/ETO SLs (this work; colored stars) is twice as large as that of the LSTO/STO SLs (previous reports; black circles). All black marks are error bars for the *S* measurements. Error bars indicate the standard error of the $\Delta V$–$\Delta T$ slope obtained from linear fitting across each single film and superlattice. They reflect the uncertainty in the measurement itself rather than variations between different samples.

## 2.4 2D confinement-induced modification of electronic structure and orbital distribution

The partial DOS (PDOS) calculated by hybrid density functional theory (DFT) calculations consistently supports the modification of electronic structure and charge distribution of Ti $3d$-orbitals within the quasi-2D layer in fractional LETO/ETO SLs. As shown in figure 4(a), the electronic structure of the fractional LETO/ETO SL exhibited Ti $3d$-electrons accumulation at slab 4 (LETO layer). The charge distribution, i.e., $|\psi|^2$, consistently demonstrates the existence of the $d$-electrons (orbitals denoted as light blue) at slab 4 (figure 4(b)). The orbital-resolved DOS further indicates that mostly Ti $3d_{xy}$-orbitals were confined near slab 4, resulting in a direct and quantitative measure of the effective potential barrier ($\Delta E_{barrier}$) height of ~0.13 eV, which resulted in 2D-like transport behaviors (figure 4(c)). Similar observation has been reported for fractional LSTO/STO SLs [57], however, LETO/ETO SL exhibited a more localized nature because of the smaller $\varepsilon$ and shorter $a_0$ of ETO than those of STO. In particular, our calculation of the LSTO/STO SL shows that the Ti $3d_{xy}$-orbitals diffused across the slabs, as evidenced by the charge distribution and orbital-resolved DOS (figure S10, Supplementary data). The localized charge profile in the LETO/ETO SL provides theoretical evidence for larger $|S_{2D}|$ and $S_{2D}/S_{3D}$. On the other hand, the $d$-electrons in the LETO single film were more or less uniformly distributed, and exhibited the same orbital characteristics as the $3d_{xy}$-orbital for both the first and second nearest neighbor (NN) regions (figure S11, Supplementary data). The LSTO single film also showed a similar distribution of $d$-electrons, qualitatively identical to that of the LETO single film (figure S12, Supplementary data). Therefore, our calculations support the enhanced 2D-ness of the fractional LETO/ETO SLs.

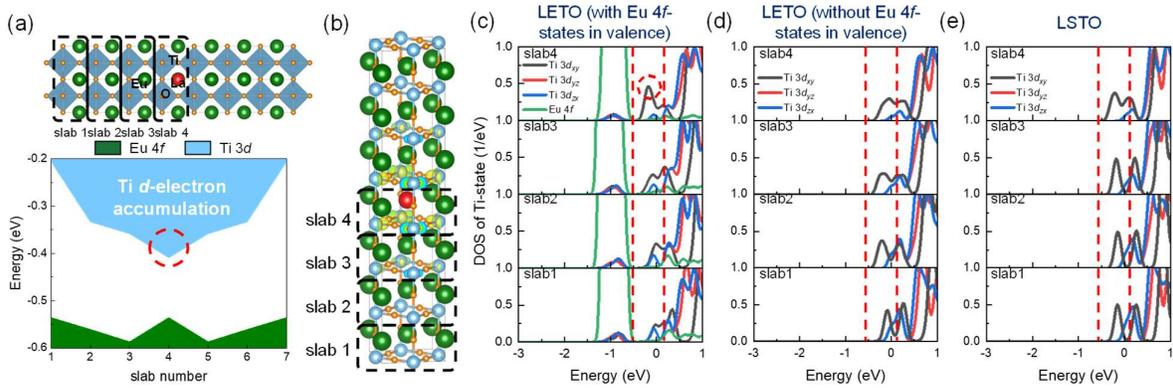

**Figure 4.** 2D confinement and orbital distribution of Ti $3d$-electrons in fractional LETO/ETO SL observed by DFT calculation. (a) Modeled structure of the fractional LETO/ETO SL and Ti $3d$-electron accumulation in the LETO layer (slab 4, marked by the red dotted circle). (b) Charge distribution near the $E_F$, represented by the light blue orbitals, showing a higher charge accumulation at slab 4 compared to the other slabs. (c-e) Layer-resolved Ti $3d$-states for fractional LETO/ETO SL with and without Eu $4f$-states, and for fractional LSTO/STO SL. At slab 4, the Ti $3d_{xy}$-orbital predominates near the $E_F$ (red dotted circle) and is closer to the Eu $4f$-orbital, suggesting a larger contribution to the Eu $4f$-Ti $3d$ interaction. Charge localization near La layer (slab 4) is enhanced (see the region between the vertical red dashed lines) when Eu $4f$-states are included, highlighting their role in reinforcing 2D confinement and potentially increasing $|S|$.

To understand the origin of the enhanced $|S|$ in fractional LETO/ETO SLs, figure 4(c)-(e) show the PDOS of Ti $3d$-states across the LETO/ETO and LSTO/STO SLs, comparing calculations with and without Eu $4f$-states in the valence. When Eu $4f$-states are included (figure 4(c)), charge carriers are more strongly localized near the La-doped layers, indicating enhanced quasi-2D confinement. This localization is reduced when the $4f$-states are treated as frozen core (figure 4(d)), resulting in a more delocalized distribution similar to the LSTO/STO case (figure 4(e)), where no $4f$-states are present. In addition, Eu $4f$-states in the valence increases the $\Delta E_{barrier}$ from 0.10 eV to approximately 0.13 eV at the La-doped layer (figure S13, Supplementary data). This $\Delta E_{barrier}$ increase is accompanied by a narrower and deeper quantum-well-like potential, indicating enhanced localization of the Ti $3d$-electrons. Consistent with an electrostatic mechanism, these results suggest that Eu $4f$-states play a key role in modifying the local potential landscape (a measurable upward shift of the local Ti $3d$ conduction edge) and strengthening the confinement potential. At the same time, the PDOS shows negligible hybridization between the Eu $4f$- and Ti $3d$-states due to an energy separation of ~1 eV. Hence, stronger confinement arises from a redistributed electrostatic potential rather than from hybridized bound states (figure 4(c)-(e)). While we performed DFT-based $m^*$ calculations for both bulk and SL of ETO- and STO-based systems (figure S14 and S15, Supplementary data), we note that these values do not capture many-body interactions such as confinement-enhanced electron–phonon coupling, which may be more relevant to the observed thermopower enhancement in LETO/ETO SLs. These findings reinforce that the enhanced |$S$| in LETO/ETO SLs originates primarily from the stronger 2D confinement induced by Eu 4*f*-states. All measurements and the theoretical evaluation refer to 300 K, since bulk ETO develops long-range antiferromagnetic order only below 4 K, spin-order mediated exchange is not expected to influence the room-temperature thermopower discussed here.

## 3. Conclusion

In this study, we explored the enhancement of |$S$| in LETO single films and fractional $\delta$-doped LETO/ETO SLs through the various parameters of dimensionality and electronic correlation. Our findings demonstrate 2D confinement of *d*-electrons within the quasi-2D layer results in a substantial enhancement of |$S$|, achieving values as high as 950 µV K$^{-1}$, twice larger than that of the LETO single films. $S_{2D}/S_{3D}$ can be achieved up to ~20 in the fractional LETO/ETO SL, which has improved about twice that of comparable fractional LSTO/STO SLs. This enhancement can be explained by the reduction of the $a_0$ and enhanced interaction between the Eu 4*f*- and Ti 3*d*-orbitals in the ETO-based systems. Hybrid DFT calculations further confirmed the confinement of the Ti 3*d*-electrons within the LETO layers, in particular, the Ti 3*d*$_{0}$-orbital is contributed to the observed |$S$| enhancement due to the presence of Eu 4*f*-orbital. Direct verification of *n* by Hall measurements on suitably designed superlattice devices, together with conductivity optimization, would enable a more quantitative evaluation of overall thermoelectric performance including *PF*. If combined with recently developed membrane fabrication techniques for freestanding complex-oxide films, our approach may provide a route toward flexible oxide-based energy-conversion platforms and enable direct in-plane $\kappa$ measurements for obtaining *ZT* in strongly confined systems [62]. Beyond its fundamental significance in paving the way for future design strategies that combine material search with 2D confinement, our study is applicable to thermal sensing, such as potentially high-efficiency sensitivity and flexible thermoelectric applications of transition metal oxide heterostructures.

## 4. Experimental Section

### 4.1 Atomic scale precision epitaxy of thin films and SLs growth

The samples were grown on LSAT substrates using atomic-scale precision growth of pulsed laser epitaxy [63]. Approximately 30–40-nm-thick epitaxial LETO single films (*x* = 0.05, 0.1, 0.2, 0.4, 0.6, 0.8, and 1) and fractional LETO/ETO SLs were fabricated on (001)-oriented LSAT substrates using pulsed laser epitaxy. Eu$_2$Ti$_2$O$_7$ and La$_2$Ti$_2$O$_7$ ceramic targets and an excimer KrF laser ($\lambda$ = 248 nm, IPEX864; Lightmachinery) with a fluence of 1.52 J cm$^{-2}$ and a repetition rate of 5 Hz were used. The thin films and superlattices were synthesized at 700 °C under an oxygen partial pressure under 4 × 10$^{-7}$ Torr. We deliberately synthesized [(LETO)$_1$|(ETO)$_{10}$]$_{10}$ SLs (*x* = 0.05, 0.1, 0.2, 0.4, 0.6 and 0.8), of which the supercells were systematically repeated 10 times along the (001)-orientation by controlling the number of laser pulses.

### 4.2 Lattice structure characterization

XRD $\theta$-2$\theta$ and XRR measurements for the LETO single films and fractional LETO/ETO SLs were performed using high-resolution XRD (PANalytical X'Pert Pro). We deduced the thicknesses of ETO and LETO layers using XRR simulations. The estimation of the superlattice peak positions using Bragg's law as $\Lambda$ = *i*$\lambda$/2 (sin$\theta_{SL}$)$^{-1}$, where $\Lambda$, *i*, $\lambda$, and $\theta_{SL}$ are the period thickness, X-ray wavelength, SL peak order, and *i*th-order SL peak, respectively.

### 4.3 Atomic structure characterization

The fractional $\delta$-doped LETO/ETO SL samples for the cross-sectional STEM observation were prepared by a dual-beam focused ion beam equipment (Helios G3, FEI, USA), which uses a Ga ion. STEM imaging and EDS mapping experiments were performed at 200 kV of accelerating voltage using JEM ARM-200F (JEOL Ltd., Japan) equipped with spherical aberration (C$_s$) corrector (ASCOR, CEOS GmbH, Germany) at Materials Imaging & Analysis Center of POSTECH in Republic of Korea. The convergence angle and probe diameter of the electron beam were about 28 mrad and 70 pm, respectively, and the collection angles for HAADF imaging were set from 54 to 216 mrad. To reduce background noises, raw STEM images were filtered using a band-pass filter (HREM Research Inc., Japan). The STEM-EDS mapping was performed using a dual-energy dispersive X-ray spectrometer (JED-2300T, JEOL Ltd., Japan) with an energy resolution of 0.01 keV. The resulting elemental maps were obtained during about 50 minutes of total acquisition time with a resolution of 512 × 512 pixels and an acquisition time of 0.01 ms per pixel. The EDS mapping results were filtered using a Wiener filter to reduce the background noises.

### 4.4 Electronic transport properties

$\sigma$, $n$, and $\mu$ were measured at room temperature through the conventional DC four-probe method, using an In-Ga alloy electrode with van der Pauw geometry. $S$ was measured at room temperature by creating a temperature difference of ~10 K across the film using two Peltier devices, while monitoring the actual temperatures at each end of the LETO single films and fractional LETO/ETO SLs using two small thermocouples. The thermo-electromotive force ($\Delta V$) and temperature difference ($\Delta T$) were measured simultaneously, and $S$ was obtained from the slope of the $\Delta V$–$\Delta T$ plots. Experimental configuration is detailed in figure S16 (Supplementary data).

### 4.5 Theoretical calculation

First-principles calculations were performed using DFT and the screened hybrid functional of Heyd–Scuseria–Ernzerhof (HSE06) [64, 65], implemented with the projector augmented-wave method in the VASP code [66, 67]. The wavefunctions were expanded in a plane-wave basis set with an energy cutoff of 400 eV. The experimental lattice constant (= 3.905 Å) of ETO at 300 K was used to match the measurement conditions, and the antiferromagnetic ordering of the Eu $4f$-spins were employed to address the correct electronic structure. To simulate fractional LETO/ETO SLs and LETO single films, $\sqrt{2} \times \sqrt{2} \times$ 7 SL and $2\sqrt{2} \times 2\sqrt{2} \times 2$ supercell were utilized, respectively. For integration over the Brillouin zone, we use a 3 × 3 × 1 $k$-point grid for the LETO/ETO SLs and a 2 × 2 × 2 $k$-point grid for the LETO supercell. The atomic positions were relaxed until the Hellmann–Feynman forces were reduced to less than 0.05 eV Å$^{-1}$.

### Data availability statement

All data that support the findings of this study are included within the article (and any supplementary files).

### Acknowledgements

This work was supported by the Basic Science Research Programs through the National Research Foundation of Korea (NRF-2021R1A2C2011340, RS-2023-00220471, and RS-2023-00281671), Grants-in-Aid for Scientific Research A (22H00253) from the Japan Society for Promotion of Science (JSPS), the National Supercomputing Center with supercomputing resources, including technical support (KSC-2024-CRE-0027), the Crossover Alliance to Create the Future with People, Intelligence and Materials, the Network Joint Research Centre for Materials and Devices, and the Korea Basic Science Institute (National Facilities and Equipment Center) grant funded by the Ministry of Education (2020R1A6C101A202).

### Conflict of interest

The authors declare no conflict of interest.

### CRediT authorship contribution statement

[#]These authors contributed equally. **Conceptualization**, D.S., M.C., and W.S.C.; **Data curation**, D.S., I.K., M.-S.K., Y.-Q.Z., W.T.L., S.-Y.C., and H.O.; **Formal analysis**, D.S., I.K., M.-S.K., Y.-Q.Z., W.T.L., S.-Y.C., and H.O.; **Investigation**, D.S., I.K., M.-S.K., and H.O.; **Validation**, D.S., S.-Y.C. H.O. and W.S.C.; **Writing – Original Draft**, D.S., I.K., S.-Y.C., M.C., H.O., and W.S.C.; **Writing – Review & Editing**, D.S., I.K., M.-S.K., S.-Y.C., M.C., H.O., and W.S.C.; **Funding Acquisition**, S.-Y.C., M.C., H.O., and W.S.C.; **Resources**, S.-Y.C.; **Supervision**, S.-Y.C., M.C., and W.S.C.

### ORCID iD

[*]**Minseok Choi** - https://orcid.org/0000-0001-8967-3030
[*]**Hiromichi Ohta** - https://orcid.org/0000-0001-7013-0343
[*]**Woo Seok Choi** - https://orcid.org/0000-0002-2872-6191

# Supplementary data

# Improving 2D-ness to enhance thermopower in oxide superlattices

**Dongwon Shin[1,#], Inseo Kim[2,#], Min-Su Kim[3], Yu-Qiao Zhang[4], Woo Tack Lim[1], Si-Young Choi[3,5,6], Minseok Choi[2,\*], Hiromichi Ohta[7,\*], and Woo Seok Choi[1,\*]**

[1] Department of Physics, Sungkyunkwan University, Suwon 16419, Republic of Korea
[2] Department of Physics, Inha University, Incheon 22212, Republic of Korea
[3] Department of Materials Science and Engineering, Pohang University of Science and Technology, Pohang 37673, Republic of Korea
[4] School of Chemistry and Chemical Engineering, Jiangsu University, Zhenjiang 212013, China
[5] Department of Semiconductor Engineering, Pohang University of Science and Technology, Pohang 37673, Republic of Korea
[6] Center of Van der Waals Quantum Solids, Institute for Basic Science (IBS), Pohang 37673, Republic of Korea a
[7] Research Institute for Electronic Science, Hokkaido University, Sapporo 001-0020, Japan

*Author to whom any correspondence should be addressed.

E-mail: hiromichi.ohta@es.hokudai.ac.jp, minseok.choi@inha.ac.kr, and choiws@skku.edu

## S1. Perovskite EuTiO₃ (ETO) and LaTiO₃ (LTO)

ETO [$Eu^{2+}Ti^{4+}(O^{2-})_3$], cubic in bulk, $a = 3.905$ Å, band insulator) and LTO [$La^{3+}Ti^{3+}(O^{2-})_3$], orthorhombic in bulk, $a = 5.634$ Å, $b = 5.616$ Å, $c = 7.915$ Å, Mott insulator) are target materials to form superlattices (SLs) [1-3]. La doping with a doping ratio of $x$ effectively modulated the Fermi level ($E_F$). Because ETO ($3d^0$) and LTO ($3d^1$) comprise the same transition metal site (Ti atom), the doping of La into ETO (LETO) nominally results in the facile introduction of the $3d$-electrons of the Ti site. Introducing La into ETO ($x > 0$) induces an insulator-to-metal transition. Upon further doping ($x$ approaching 1.0), the conduction band of the Ti $3d$-orbital splits into the upper (UHB) and lower Hubbard bands (LHB), resulting in a Mott insulating phase.

## S2. Structural properties of La$_x$Eu$_{1-x}$TiO₃ (LETO) single films and fractional $\delta$-doped LETO/ETO SLs

X-ray diffraction (XRD) $\theta$–$2\theta$ scans of the LETO single films with distinct fringe on $(LaAlO_3)_{0.3}(Sr_2TaAlO_6)_{0.7}$ (LSAT) substrates show the Bragg diffraction peaks of the high-quality LETO thin films (002) plane at ~45.0° and those of the LSAT substrates (002) plane at 46.9° in figure S2(a). The LETO single films ($x < 1.0$) were fully strained to the LSAT substrates, as evidenced by XRD reciprocal space mappings (RSMs) near the (103) plane of the LSAT substrate (figure S2(b)). Only the $x = 1.0$ (LTO) single film showed signs of strain relaxation, owing to the largest lattice mismatch of 2.28% [4, 5]. In fractional LETO/ETO SLs, both the XRD and X-ray reflectivity (XRR) curves demonstrate SL Bragg peaks (figure 2(e) and 2(f)) and fringes, verifying the high quality of the SLs. Note that the coherent SL Bragg peaks are rather subtle because of the similar material densities of ETO (6.62 g/cm³) and LTO (6.28 g/cm³), which becomes further indistinguishable with small $x$. Nevertheless, the SL Bragg peak positions were derived from Bragg's law, $\Lambda_{SL} = i\lambda/2 \times (\sin\theta_{SL})^{-1}$, where $\Lambda_{SL}$, $i$, $\lambda$, and $\theta_{SL}$ are the period thickness, SL peak order, X-ray wavelength, and $i$th-order SL peak angle, respectively, consistent with the experimental results in figure S3(a). Figure S3(b) shows the RSMs near LSAT (103), indicating that the fractional LETO/ETO SLs are fully strained.

## S3. Microstructure of fractional $\delta$-doped LETO/ETO SLs

Coherent alignment of the atomic columns without defects and/or misfit dislocations at the interfaces is shown. The stripe-shaped contrast is observed in the $Z$-contrast high-angle annular dark-field scanning transmission electron microscopy (HAADF-STEM) image, where the brighter region is attributed to the ETO layer with slightly heavier Eu ions (atomic number, $Z = 63$), and the darker region is attributed to the LETO layer doped with lighter La ions ($Z = 57$). The energy-dispersive X-ray spectroscopy (EDS) map showed potential intermixing between the Eu and La ions. Despite the potential intermixing, the total supercell thickness (11 u.c.) of the SLs was consistently fixed at $x = 0.1$ (figure 2(g) and 2(h)), 0.6, and 1 (figure S4 and S5) SLs, respectively.

## S4. Transport measurements of LETO single films and fractional LETO/ETO SLs

Both LETO single films and fractional LETO/ETO SLs exhibit $n$-type conduction owing to the negative slope of $\Delta V$-$\Delta T$ curves in figure S6. The LETO single film at $x = 0.8$ exhibits unexpected behavior with a slight enhancement of $|S_{3D}|$ (see the orange open circle in figure 3(b)), accompanied by $S_{2D}/S_{3D}$ suddenly dropped at $x = 0.8$, as shown in figure 3(c). To further elucidate this trend, we performed electrical transport measurements of the LETO single films, as shown in figure S7. Both the electrical conductivity and carrier density ($\sigma$ and $n$) display an overall increasing trend up to $x = 0.8$ as shown in figure S7(a) and 7(b), similar to epitaxial La-doped SrTiO₃ (LSTO) thin films [6]. We calculated the power factor ($PF = S^2\sigma$) of LETO single films, as shown in figure S7(c), which indicates a similar behavior to $|S_{3D}|$. In addition, as shown in figure S7(d), the mobility ($\mu$) increases from 0.25 to 2.00 cm² V⁻¹ s⁻¹ as $x$ increases from 0.05 to 0.4. Above $x = 0.4$, $\mu$ decreases from 2.00 to 1.00 cm² V⁻¹ s⁻¹. However, these trends do not fully explain the noticeable discontinuity of $|S_{3D}|$ in the LETO single films and $S_{2D}/S_{3D}$ at $x = 0.8$. Therefore, the effective mass was investigated to understand this anomaly. In contrast to the LETO single films, the transport signal of fractional LETO/ETO SLs was below the reliable limit of our measurement system due to the very low carrier concentration $n$ and mobility $\mu$. We could only measure conductivity $\sigma$ of fractional LETO/ETO SLs, shown in figure S8.

## S5. Unexpected behavior of $|S_{3D}|$ and $S_{2D}/S_{3D}$

The enhancement of the effective mass ($m^*_{DOS}$) of the density of state (DOS) in the Ti $3d$-orbital can explain the noticeable discontinuity of $|S|$ in the LETO single films and $S_{2D}/S_{3D}$ at $x = 0.8$ in figure S7(e). Based on the electron effective calculations, an increase in the effective mass $m^*$ ($m^* = m^*_{DOS}/$ #. of $t_{2g}$ electrons with spin-up and -down: $3 \times 2 = 6$) was confirmed, which resulted in the unexpected behavior of $|S_{3D}|$ in the epitaxial LETO single film at $x = 0.8$ owing to the Mott transition boundary [6]. The detailed expression of $m^*$ is explained by an electrical model for the electron-phonon scattering [7]. To verify the enhancement in $m^*$, we determined the $m^*_{DOS}$, using $S$ and $n$ according to the following Equation S1–S4.

$$S = \frac{k_B}{e}\left[\frac{(1+s+r)F_{s+r}(\xi)}{(s+r)F_{s+r-1}(\xi)} - \xi\right]$$ (Equation S1)

where $k_B$, $\xi$, $r$, and $F_r$ are the Boltzmann constant, chemical potential, scattering parameter of relaxation time, and Fermi integral, respectively. Here, $s = 1$ is for 3D system, while $s = 1/2$ is for 2D system [8]. In our case, the epitaxial films follow 3D behaviors. Therefore, the equation can be simplified as:

$$S = \frac{k_B}{e}\left[\frac{(2+r)F_{r+1}(\xi)}{(1+r)F_r(\xi)} - \xi\right]$$ (Equation S2)

where $r = 0$ for acoustic phonon scattering, $r = 1/2$ for polar optical phonon scattering, and $r = 1$ for nonpolar optical phonon scattering.

$$F_r(\xi) = \int_0^\infty \frac{x^r}{1 + e^{x-\xi}}dx$$ (Equation S3)

and $n$ is described by

$$n = 4\pi\left(\frac{2m^*_{DOS}k_B T}{h^2}\right)^{\frac{3}{2}}F_{\frac{1}{2}}(\xi)$$ (Equation S4)

where $h$ and $T$ are Planck's constant and absolute temperature, respectively. $r$ can be defined as $1/2$, because the carriers can be scattered by polar optical phonons. This selection is consistent with prior reports on oxide thermoelectric systems. At $T = 300$ K, polar optical phonon scattering is widely accepted as the dominant carrier scattering mechanism in perovskite oxides such as STO and its analogs [6, 9]. In our case, ETO shares an almost identical lattice structure and similar phonon dispersion with STO, with the only difference being the substitution of Sr by Eu. In case of ETO, the electron scattering was reported to be similar with that of STO, which is due to the polar optical phonons [10]. Therefore, $r = 1/2$ has been selected in the calculation. These factors reinforce the assumption that similar polar optical phonon scattering is dominant in our ETO-based superlattices at 300 K. As shown in figure S7(e), $m^*$ gradually decreases up to $x = 0.6$, in contrast, $m^*$ suddenly increases from 0.59 to 0.78 at $x = 0.8$, because the electronic band structure of LETO thin films exhibits Mott boundary across $x = 0.8$. The $n$-type conduction of the LETO single film $x = 0.8$ is still shown because the LETO single film at $x = 0.8$ might be attributed to an incomplete opening of the Hubbard band owing to the structural deformation induced by the LSAT substrate [11, 12]. Therefore, the Mott transition boundary near $x = 0.8$ exhibits an enhancement of $m^*$ and a strong correlation attributed to the larger $|S|$ in LETO single films and lower $S_{2D}/S_{3D}$ owing to the modification of the electronic structure, respectively. Notably, $m^*$ at $x = 0.05$ is lower than that at $x = 0.1, 0.2, 0.4$ and $0.6$. Because the La doping ratio is too low, $E_F$ is not well defined, and $m^*$ at $x = 0.05$ is underestimated (See Table S1).

## S6. Various $x$ dependence on $|S_{2D}|$ and $|S_{3D}|$

Fractional LETO/ETO SLs exhibit a more pronounced 2D nature with a smaller effective Bohr radius, leading to a significant change of $|S_{2D}|$ with various $x$, compared to fractional LSTO/STO SLs (figure S9). In figure S10, the charge distribution of the fractional LSTO/STO SLs exhibits less variation among the slabs compared to the fractional LETO/ETO SLs (figure 4(b) and 4(c)). In contrast, both LETO and LSTO single films (3D cases; figure S11 and S12), where La dopant can be randomly distributed, show a smaller difference in $|S_{3D}|$ between the LETO and LSTO films (figure S9). This is due to their similar electronic band structures, specifically the Ti $3d$-orbitals, as shown in Figure 4a. In addition, at a higher $x$ in fractional LETO/ETO (figure 3(b) at $x = 0.8$) and LSTO/STO SLs (figure S9 at $x = 0.75$), the $|S_{2D}|$ values are similar, as the contribution of Eu $4f$-orbital becomes weaker and negligible as the $E_F$ shifts to a higher energy level.

## Figures and Tables

**Table S1.** Reported thermopower values |*S*| (300 K) of representative oxide thermoelectric materials classified by dimensionality, including this work.

| Materials | Dimension | $|S|$ ($\mu V\,K^{-1}$) | Reference |
|---|---|---|---|
| Al-doped ZnO | 3D | < 150 | *J. Appl. Phys.* **1996,** 79, 1816-1818. |
| Electron-doped TiO$_2$ | 3D | < 110 | *J. Appl. Phys.* **2006**, 100, 096105. |
| Na$_x$CoO$_2$ single crystal | 3D | 100 | *Phys. Rev. B* **1997**, 56, R12685(R). |
| La, Dy, Yb and Y-doped CaMnO$_3$ single crystal | 3D | 40 – 135 | *Chem. Mater.* **2009**, 21, 4653–4660. |
| LaCoO$_3$ | 3D | 710 | *Appl. Phys. Lett.* **2004**, 84, 1099–1101. |
| LaTiO$_3$ thin films | 3D | < 62 | *Adv. Sci.* **2021**, 8, 2102097. |
| EuTiO$_3$ polycrystal | 3D | ∼ 1000 | *Appl. Phys. Lett.* **2012**, 101, 033908. |
| SrTi$_{1-x}$Nb$_x$O$_3$ thin films | 3D | 100 – 150 | *J. Appl. Phys.* **2019**, 126, 075104. |
| SrTi$_{1-x}$Nb$_x$O$_3$/SrTiO$_3$ superlattices | 2D | < 500 | *Nat. Mater.* **2007**, 6, 129–134. |
| La$_{1-x}$Sr$_x$TiO$_3$ thin films | 3D | 20 – 400 | *J. Appl. Phys.* **2019**, 126, 075104. |
| La$_{1-x}$Sr$_x$TiO$_3$/SrTiO$_3$ superlattices | 2D | 160 – 407 | *Adv. Mater.* **2014**, 26, 6701–6705. |
| **La$_{1-x}$Eu$_x$TiO$_3$ films** | **3D** | **18 – 443** | **This work** |
| **La$_{1-x}$Eu$_x$TiO$_3$/EuTiO$_3$ superlattices** | **2D** | **229 – 907** | **This work** |

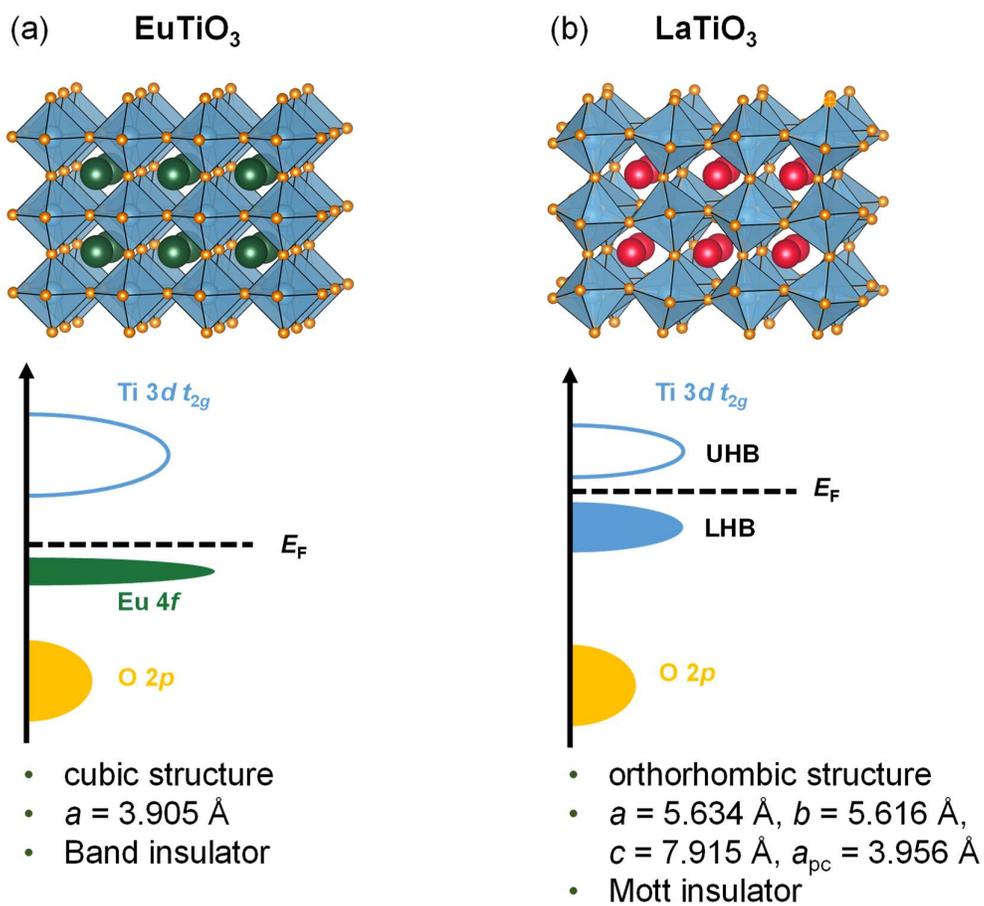

**Figure S1.** Schematic crystal and electronic structures for (a) ETO (*x* = 0) and (b) LTO (*x* = 1.0), including the structural type, lattice parameter, and simplified electronic band structure.

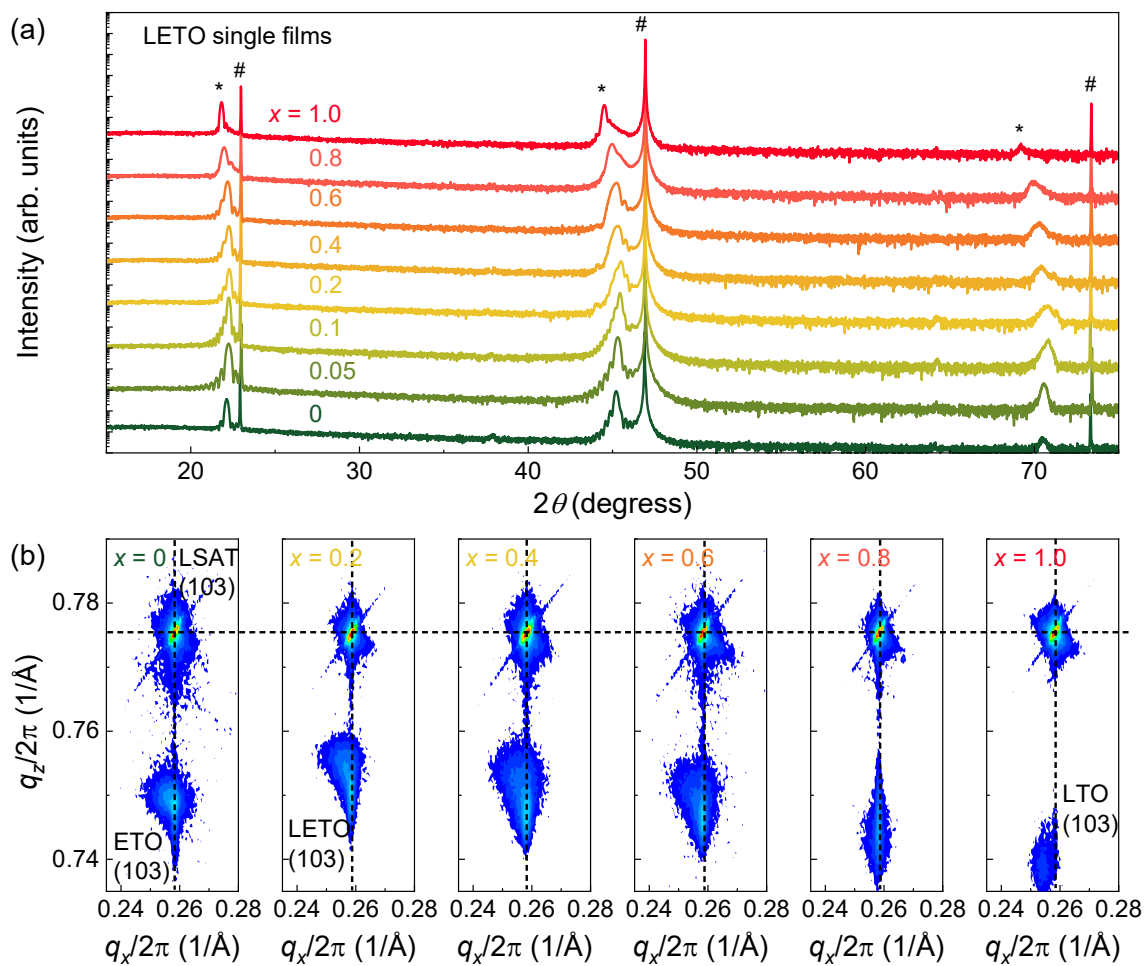

**Figure S2.** Structural characterization of epitaxial LETO single films. (a) Wide-range XRD $\theta$-$2\theta$ scan and (b) RSMs of LETO single films ($x$ = 0, 0.2, 0.4, 0.6, 0.8 and 1.0) near the LSAT (103) plane confirms the fully strained state, except for $x$ = 1.0. The horizontal and vertical lines are a guide to the eye.

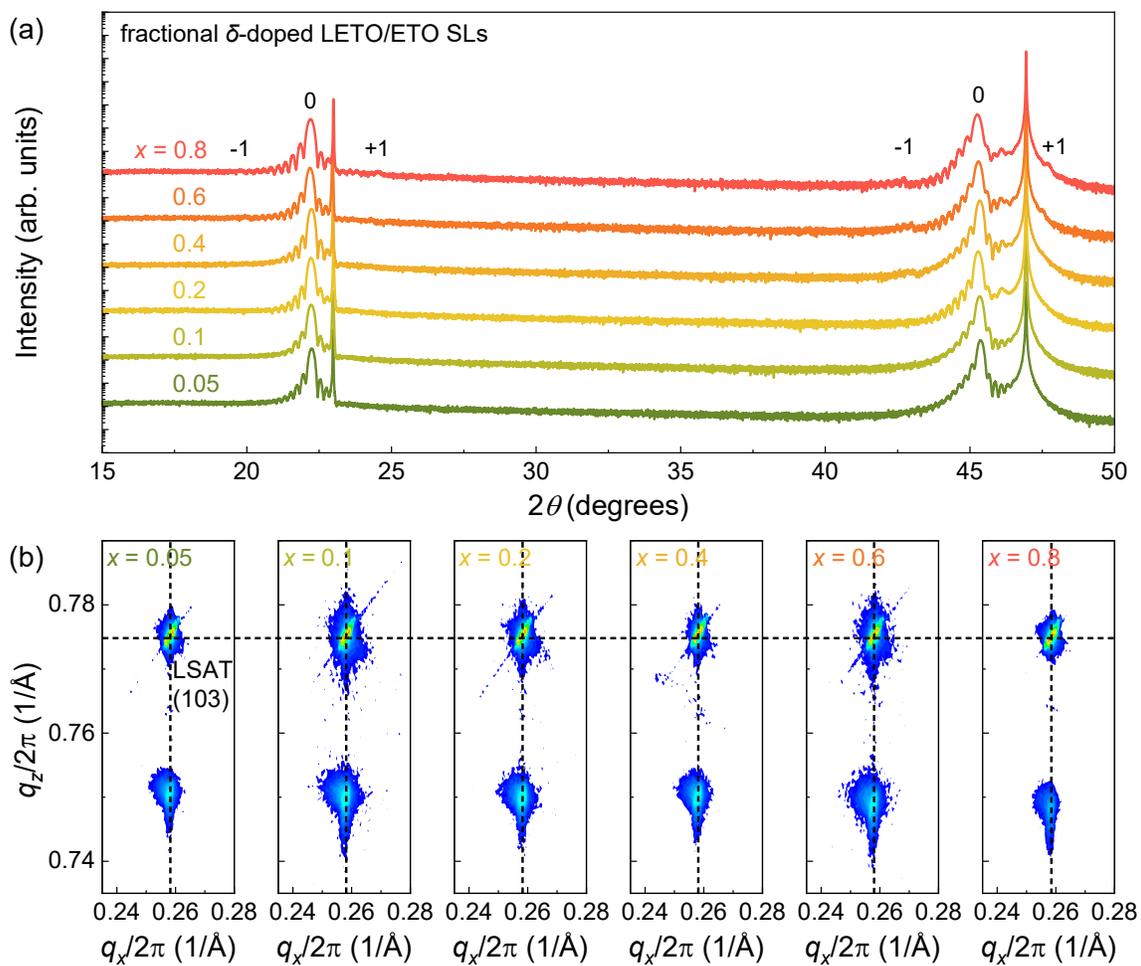

**Figure S3.** Structural characterization of atomically designed fractional LETO/ETO SLs. (a) Wide-range XRD $\theta$-$2\theta$ scan and (b) RSMs of fractional LETO/ETO SLs ($x$ = 0.05, 0.1, 0.2, 0.4, 0.6 and 0.8) near the LSAT (103) plane confirms the fully strained state. The horizontal and vertical lines are a guide to the eye.

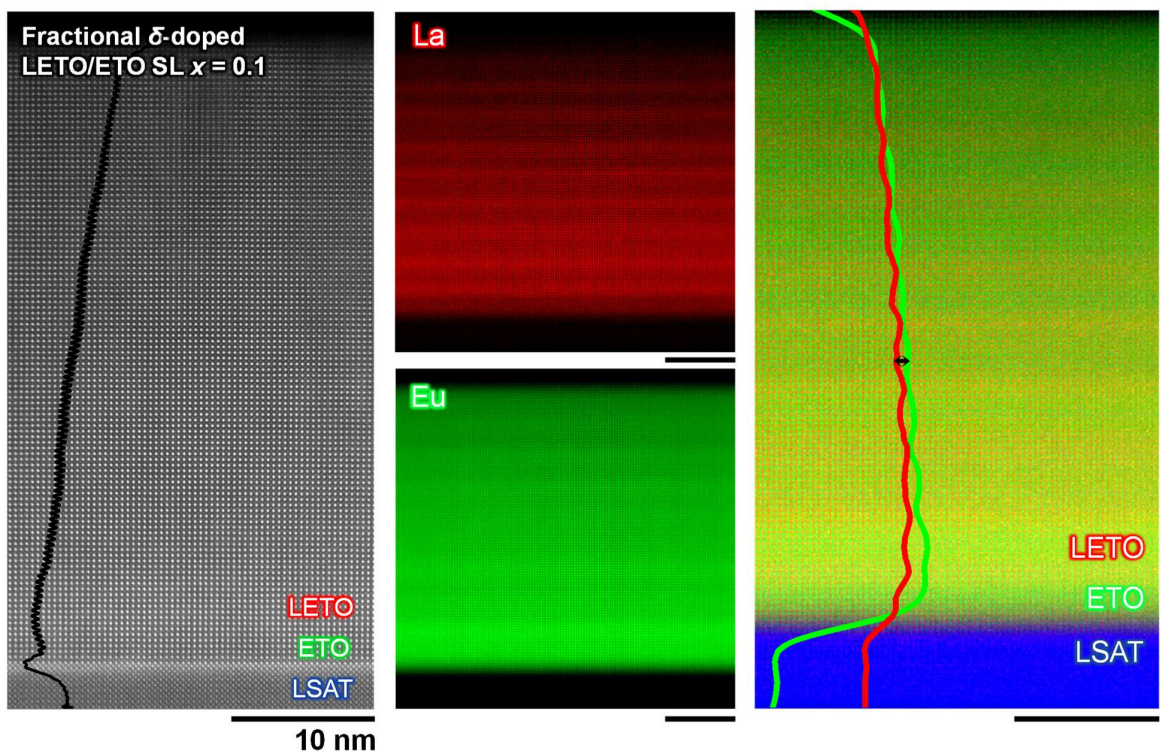

**Figure S4.** Microscopic structures of atomically designed fractional LETO/ETO SL with *x* = 0.1. The intensity and cross-sectional *Z*-contrast HAADF-STEM images for *x* = 0.1 fractional LETO/ETO SL. The scale (black bars) is 10 nm in all shown images with the atomic arrangement of the SL. The chemical formulas of the component materials in the SLs correspond to the atomic-resolution EDS image with La (red), Eu (green) and Ta (blue). The intensity difference between Eu and La atoms additionally confirmed the stacking of SL.

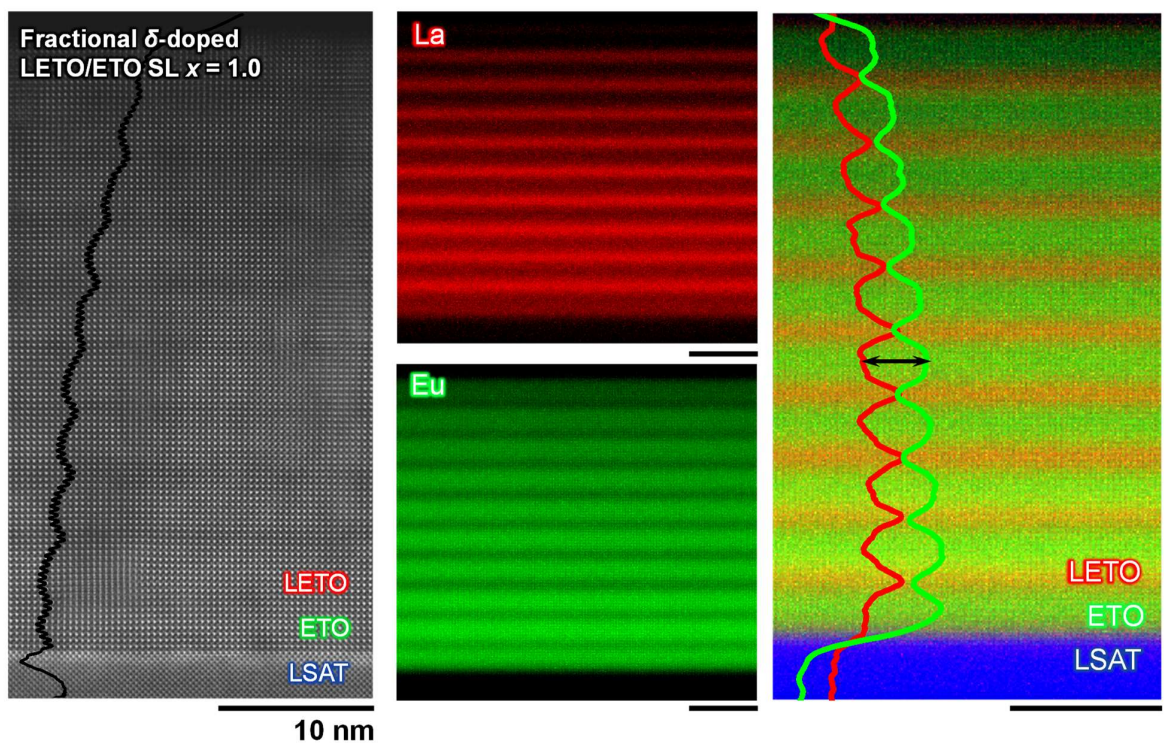

**Figure S5.** Microscopic structures of atomically designed fractional LETO/ETO SL with $x = 1.0$. The intensity and cross-sectional $Z$-contrast HAADF-STEM images for $x = 1.0$ fractional LETO/ETO SL. The scale (black bars) in all the shown images with the atomic arrangement of the SL is 10 nm. The chemical formulas of the component materials in the SLs correspond to the atomic-resolution EDS image with La (red), Eu (green) and Ta (blue). The intensity difference between Eu and La atoms at $x = 1.0$ is obviously larger than that at $x = 0.1$.

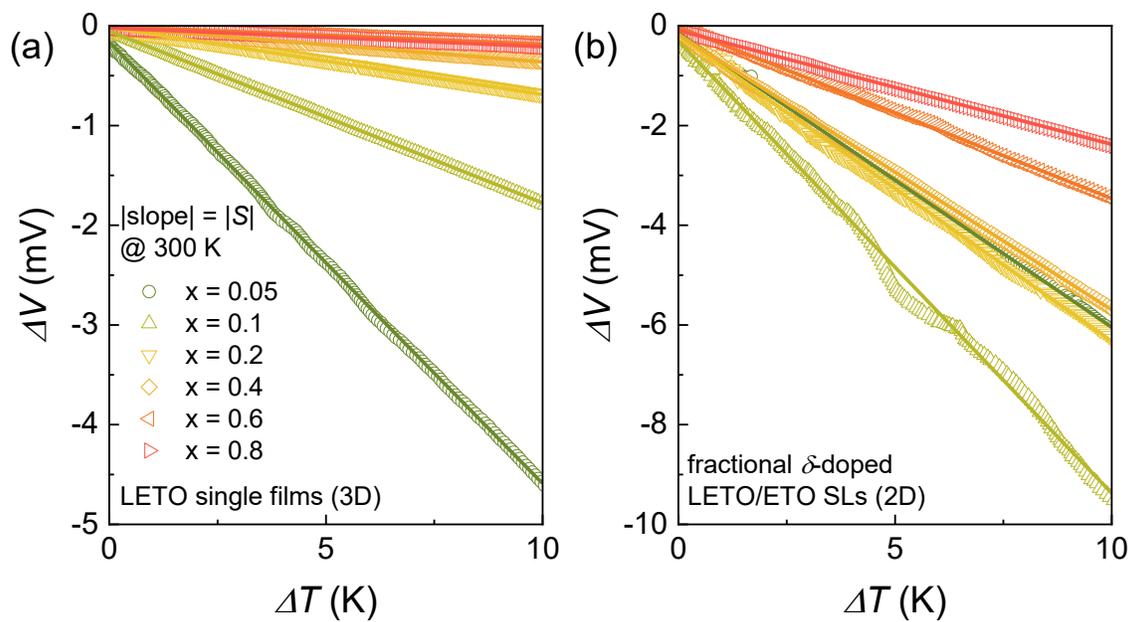

**Figure S6.** The thermo-electromotive force ($\Delta V$) as a function of temperature difference ($\Delta T$), extracting slope as $|S|$ at 300 K. (a) Epitaxial LETO single films (3D case) and (b) Fractional LETO/ETO SLs (2D case).

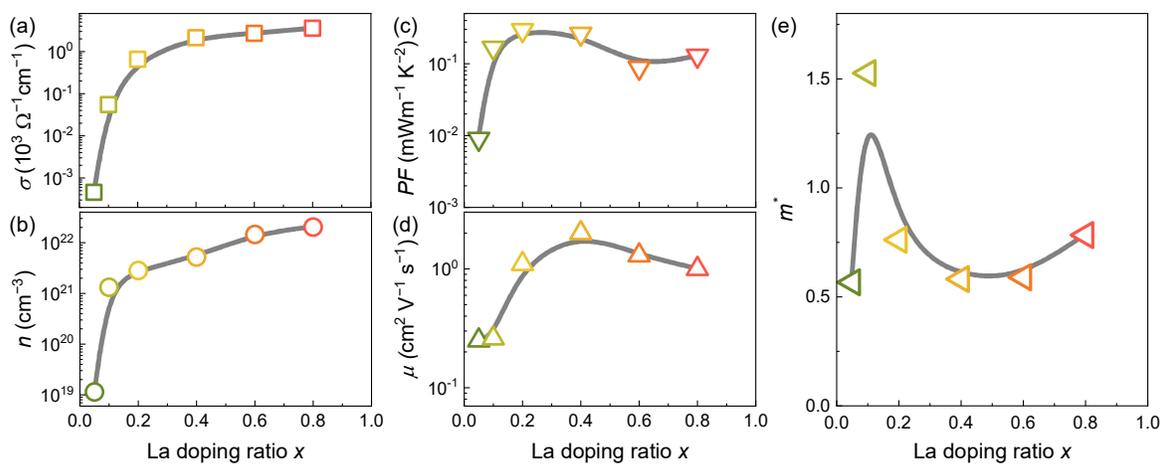

**Figure S7.** Electron transport characteristics of epitaxial LETO single films at room temperature. (a) electrical conductivity, $\sigma$, (b) carrier concentration, $n$, (c) power factor, $PF$, (d) Hall mobility, $\mu$, and (e) the effective mass, $m^*$.

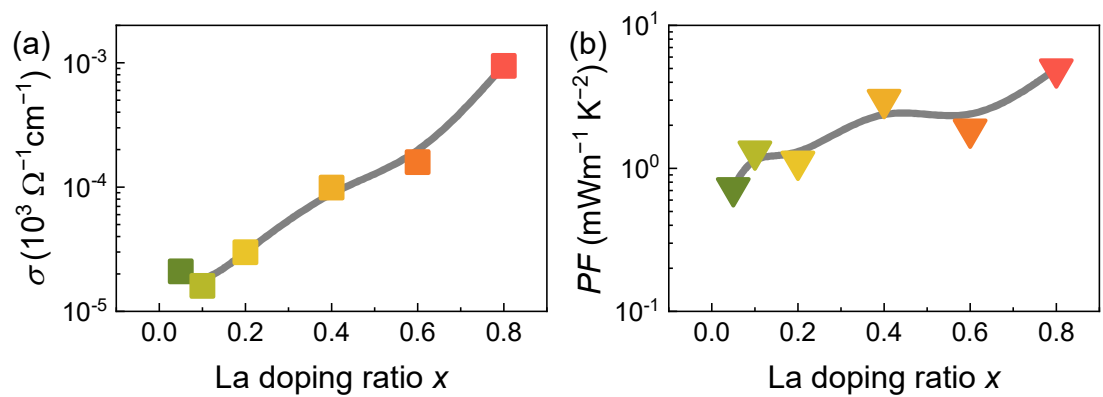

**Figure S8.** Electron transport characteristics of fractional LETO/ETO SLs at room temperature. (a) electrical conductivity, $\sigma$ and (b) power factor, *PF*.

**Table S2.** Summary of *x*-dependent electron transport characteristics of epitaxial LETO single films at room temperature. Hall mobility, $\mu$, DOS effective mass, $m^*_{DOS}$, effective mass, $m^*$, relaxation time, $\tau_s$, chemical potential, $\eta^*$, Fermi energy, $E_F$, Fermi velocity, $v_F$, and mean free path, $l_{MFP}$ are listed.

| La doping ratio *x* | $\mu$ [cm$^2$ V$^{-1}$ s$^{-1}$] | $m^*_{DOS}$ | $m^* = m^*_{DOS}/6$ | $\tau$ [s] = $\mu \cdot m^*/e$ | $\eta^*$ | $E_F$ [J] | $v_F$ [m/s] | $l_{MFP}$ [m] |
|---|---|---|---|---|---|---|---|---|
| 0 | ---- | ---- | ---- | ---- | ---- | ---- | ---- | ---- |
| 0.05 | 0.3 | 3.4 | 0.6 | 8.05E-17 | ---- | ---- | ---- | ---- |
| 0.1 | 0.3 | 9.2 | 1.5 | 2.26E-16 | 1.3 | 5.30E-21 | 87318 | 1.97E-11 |
| 0.2 | 1.1 | 4.6 | 0.8 | 4.76E-16 | 6.0 | 2.49E-20 | 267647 | 1.27E-10 |
| 0.4 | 2.0 | 3.5 | 0.6 | 6.61E-16 | 12.0 | 4.97E-20 | 433134 | 2.86E-10 |
| 0.6 | 1.3 | 3.5 | 0.6 | 4.36E-16 | 23.5 | 9.73E-20 | 601834 | 2.62E-10 |
| 0.8 | 1.0 | 4.7 | 0.8 | 4.45E-16 | 22.3 | 9.24E-20 | 508801 | 2.27E-10 |
| 1 | 0.1 | 1.9 | 0.3 | 1.59E-17 | 30.0 | 1.24E-19 | 935587 | 1.49E-11 |

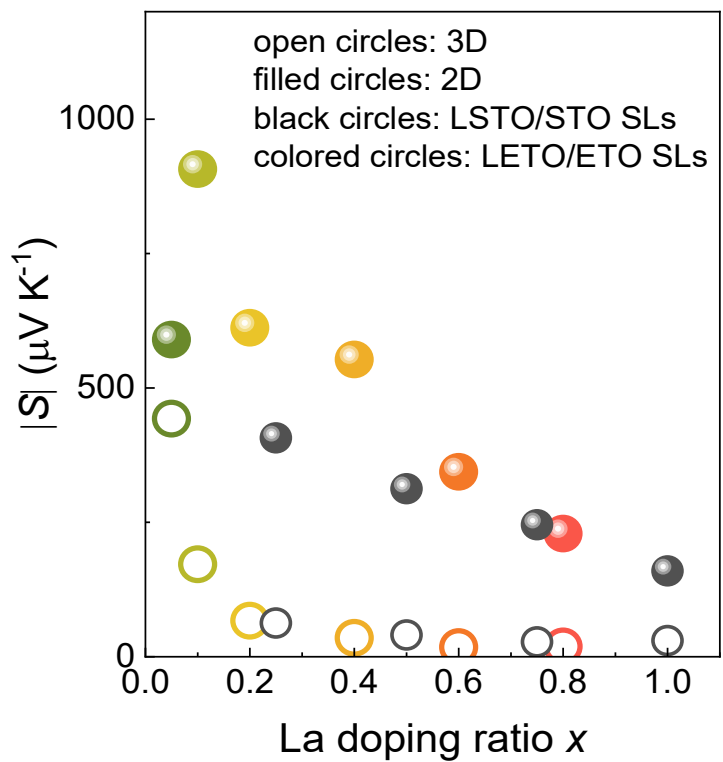

**Figure S9.** Comparison of the $S_{2D}$ and $S_{3D}$ STO- and ETO-based systems. Extracted |$S$| as a function of *x* for the LSTO single films (black open circles), fractional LSTO/STO SLs (black filled circles), LETO single films (colored opened circles) and fractional LETO/ETO SLs (colored filled circles).

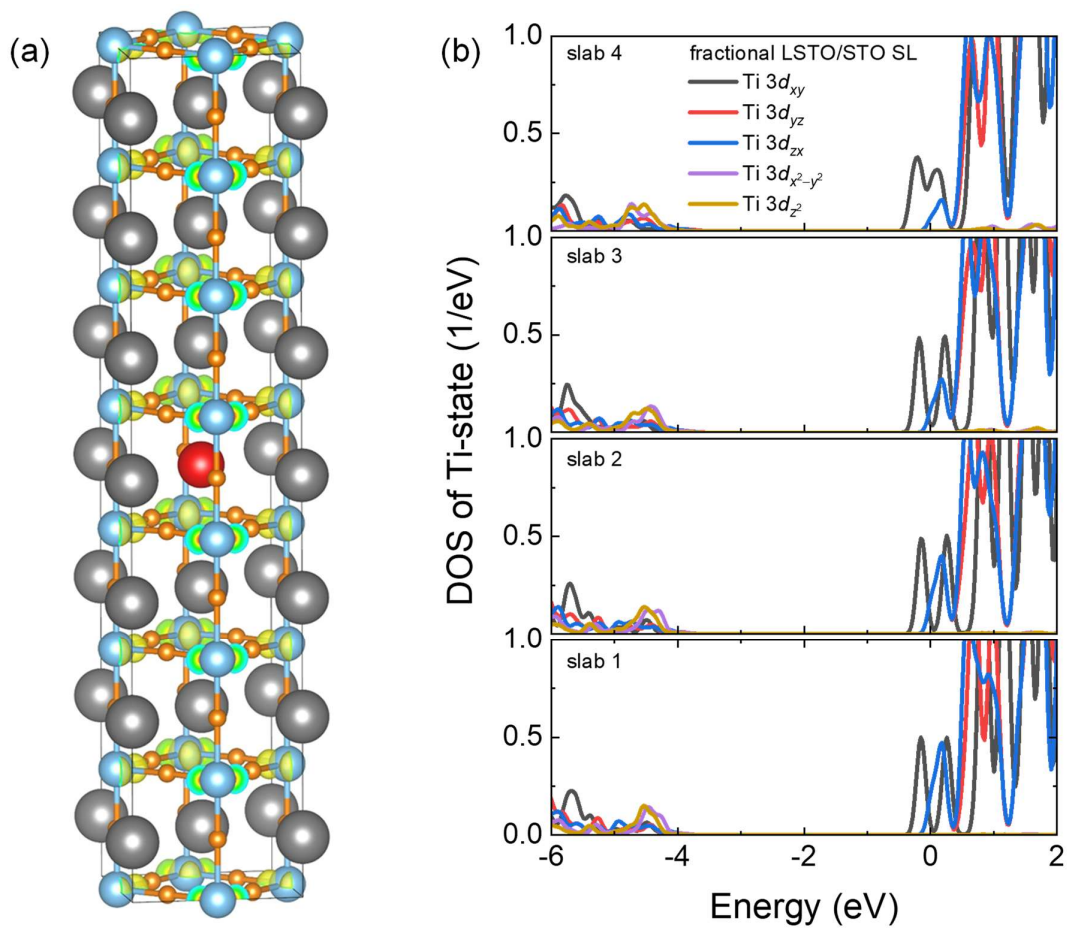

**Figure S10.** Theoretical calculation for Ti 3*d*-electrons in fractional LSTO/STO SL. (a) Charge distributions among the slabs show similar charge accumulation, compared to the fractional LETO/ETO SL. (b) Partial DOS for the Ti 3*d*-orbital in the fractional LSTO/STO SL at each slab.

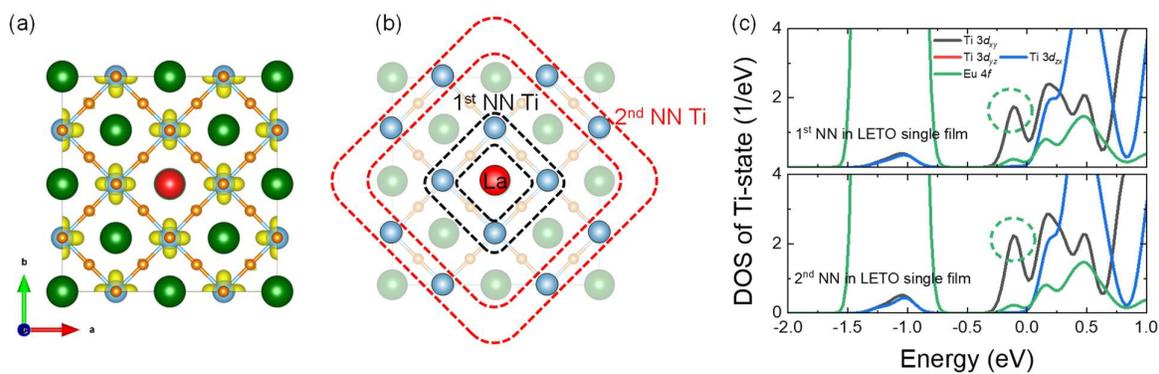

**Figure S11.** Theoretical calculations for the randomly distributed La dopant in LETO single film. (a) The charge from the La dopant is almost evenly distributed across each Ti site. (b) Definition of the 1$^{st}$ and 2$^{nd}$ NNs for Ti sites. (c) The partial DOS for the Ti 3$d$ orbitals of the 1$^{st}$ and 2$^{nd}$ NNs appears to be identical.

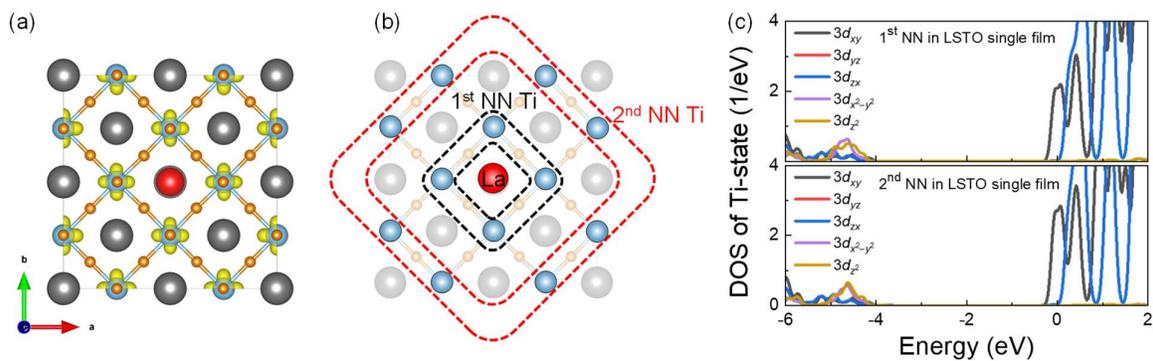

**Figure S12.** Theoretical calculations for the randomly distributed La dopant in LSTO single film. (a) The charge from the La dopant is almost evenly distributed across each Ti site. (b) Definition of the 1st and 2nd NNs for Ti sites. (c) The partial DOS for the Ti 3*d* orbitals of the 1st and 2nd NNs appears to be identical, which is similar to the LETO single film.

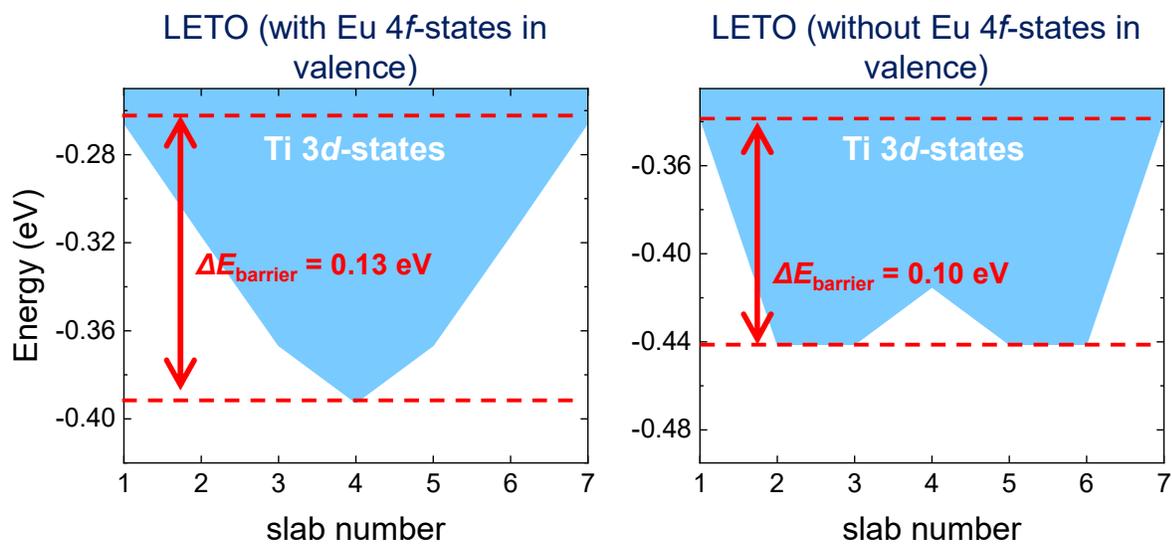

**Figure S13.** Layer-resolved Ti 3*d*-states for fractional LETO/ETO SL with and without Eu 4*f*-states. Examining the conduction-band edges of the lowest unoccupied Ti 3*d*-states reveals that including the Eu 4*f*-states in the valence increases the effective potential barrier height ($\Delta E_{barrier}$) from 0.10 eV to approximately 0.13 eV at the La-doped layer.

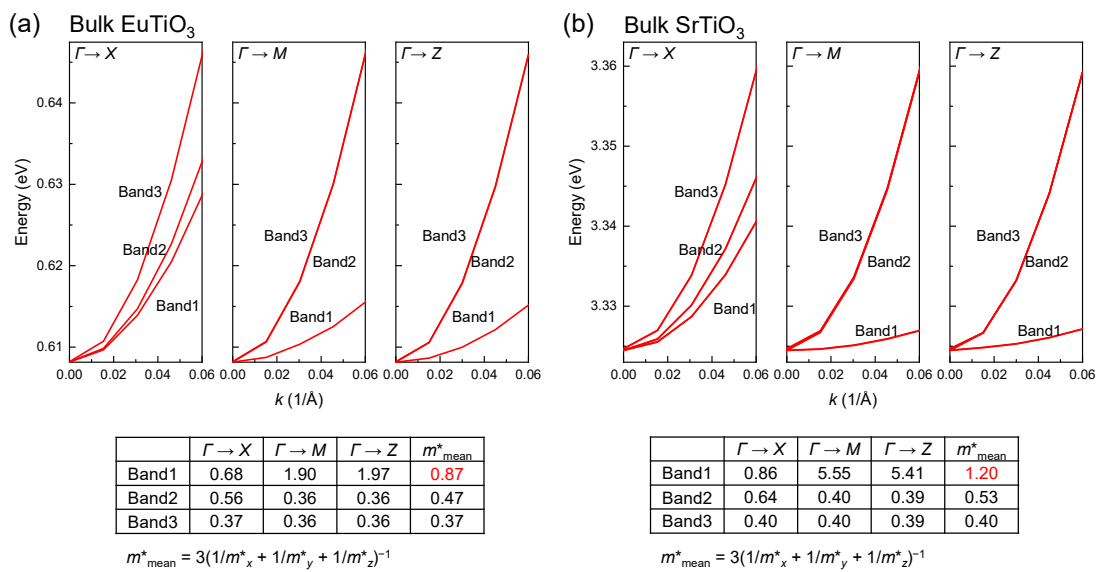

**Figure S14.** Electronic band structures and calculated $m^*$ of bulk ETO and STO from DFT calculations.

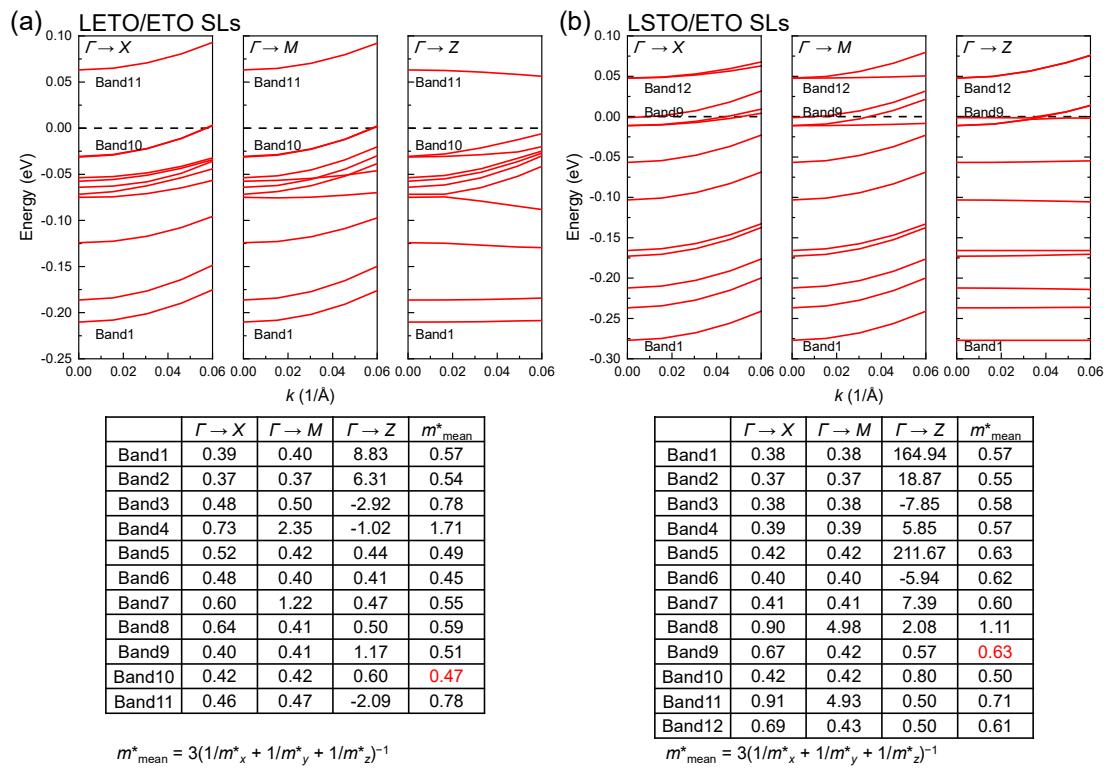

**(a)** LETO/ETO SLs

| | $\Gamma \to X$ | $\Gamma \to M$ | $\Gamma \to Z$ | $m^*_{mean}$ |
|---|---|---|---|---|
| Band1 | 0.39 | 0.40 | 8.83 | 0.57 |
| Band2 | 0.37 | 0.37 | 6.31 | 0.54 |
| Band3 | 0.48 | 0.50 | -2.92 | 0.78 |
| Band4 | 0.73 | 2.35 | -1.02 | 1.71 |
| Band5 | 0.52 | 0.42 | 0.44 | 0.49 |
| Band6 | 0.48 | 0.40 | 0.41 | 0.45 |
| Band7 | 0.60 | 1.22 | 0.47 | 0.55 |
| Band8 | 0.64 | 0.41 | 0.50 | 0.59 |
| Band9 | 0.40 | 0.41 | 1.17 | 0.51 |
| Band10 | 0.42 | 0.42 | 0.60 | 0.47 |
| Band11 | 0.46 | 0.47 | -2.09 | 0.78 |

$m^*_{mean} = 3(1/m^*_x + 1/m^*_y + 1/m^*_z)^{-1}$

**(b)** LSTO/ETO SLs

| | $\Gamma \to X$ | $\Gamma \to M$ | $\Gamma \to Z$ | $m^*_{mean}$ |
|---|---|---|---|---|
| Band1 | 0.38 | 0.38 | 164.94 | 0.57 |
| Band2 | 0.37 | 0.37 | 18.87 | 0.55 |
| Band3 | 0.38 | 0.38 | -7.85 | 0.58 |
| Band4 | 0.39 | 0.39 | 5.85 | 0.57 |
| Band5 | 0.42 | 0.42 | 211.67 | 0.63 |
| Band6 | 0.40 | 0.40 | -5.94 | 0.62 |
| Band7 | 0.41 | 0.41 | 7.39 | 0.60 |
| Band8 | 0.90 | 4.98 | 2.08 | 1.11 |
| Band9 | 0.67 | 0.42 | 0.57 | 0.63 |
| Band10 | 0.42 | 0.42 | 0.80 | 0.50 |
| Band11 | 0.91 | 4.93 | 0.50 | 0.71 |
| Band12 | 0.69 | 0.43 | 0.50 | 0.61 |

$m^*_{mean} = 3(1/m^*_x + 1/m^*_y + 1/m^*_z)^{-1}$

**Figure S15.** Electronic band structures and calculated $m^*$ of fractional LETO/ETO and LSTO/STO SLs with $x = 0.071$ from DFT calculations. $m^*$ values in both SLs are reduced compared to bulk, indicating the observed $|S|$ enhancement is not captured by band structure alone and may involve confinement-induced many-body effects.

To temperature monitors (Δ*T* measurement)

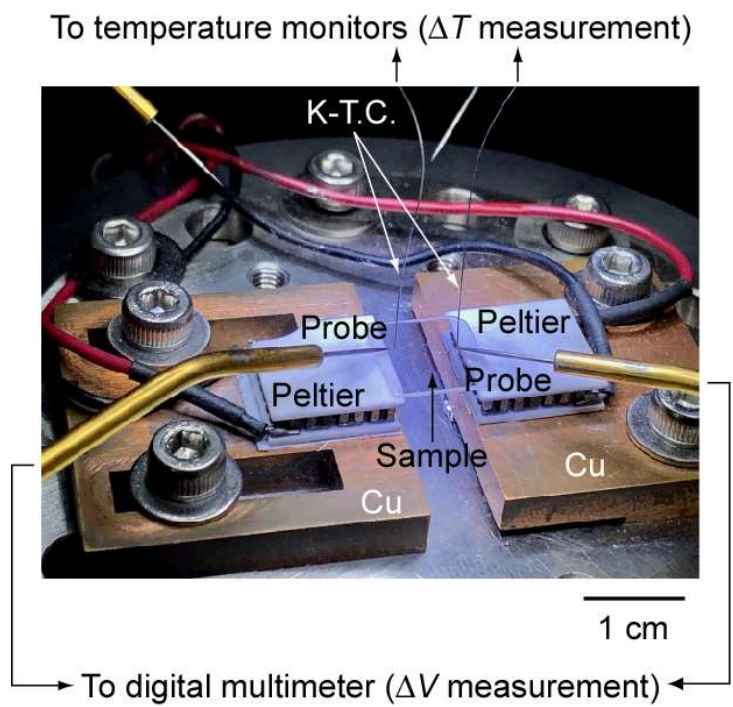

**Figure S16.** |*S*| measurement setup. The samples were put between two Peltier devices (one will be hot and the other one will be cold) to add temperature differences between two edges of the sample. The temperature difference (Δ*T*) was controlled by changing the flowing current and measured using K-type thermocouples (K-T.C.). The thermo-electromotive force (Δ*V*) was measured using a digital multimeter. The linear slope of Δ*T*−Δ*V* gave |*S*|.